\documentclass[twocolumn,showpacs,prb]{revtex4}
\usepackage{amssymb}

\usepackage{graphics}
\usepackage{epsfig}

\usepackage{dcolumn}
\usepackage{amsmath}
\usepackage{color}

\newcommand{\beq}{\begin{equation}}
\newcommand{\eeq}{\end{equation}}
\newcommand{\bea}{\begin{eqnarray}}
\newcommand{\eea}{\end{eqnarray}}
\newcommand{\bwt}{\begin{widetext}}
\newcommand{\ewt}{\end{widetext}}

\begin{document}

\title{Graphite in the bi-layer regime: in-plane transport}
\date{\today}
\author{D. B. Gutman,$^{1,2}$ S. Tongay,$^{3}$ H. K. Pal,$^{3}$ D. L. Maslov,$^{3}$
and A. F. Hebard$^{3}$}

\begin{abstract}
An interplay between the increase in the number of carriers and the decrease in the scattering time is expected to result in a saturation of the in-plane resistivity, $\rho_{ab}$, in graphite above  room temperature. Contrary to this expectation, we observe
a pronounced increase in $\rho_{ab}$ in the interval between $300$ and $900$ K.  We provide a theory of this effect based on intervalley scattering
of charge carriers by  high-frequency, graphene-like optical phonons.
\end{abstract}
% governed by competition between growing with
% temperature carrier density and scattering rate. For in plane transport the
% scattering processes win,
\affiliation{
$^1$Institut f\"ur Theorie der kondensierten
Materie, Universit\"at Karlsruhe, 76128 Karlsruhe,
Germany\\
$^2$DFG--Center for Functional Nanostructures,
Universit\"at Karlsruhe, 76128 Karlsruhe, Germany\\
$^3$ Department
of Physics, University of Florida,
Gainesville, FL 32611, USA}

\maketitle

\section{Introduction}

The family of graphite allotropes includes fullerenes, carbon nanotubes,
graphene, graphite, and diamond. A combination of chemical simplicity and
diverse physical properties, characteristic for these materials, makes
carbon-based electronics a promising field of research. Surprisingly, many
properties of the most common member of this group, graphite--such as
non-metallic $c$-axis transport \cite{graphite} and apparent quantization of
Hall resistance \cite{Hall_kopelevich}--are yet to be explained. In this
work, we focus on  in-plane transport in graphite at zero magnetic field,
which is well understood in the degenerate regime, i.e., for $T<E_F\sim 250$ K (we set $k_{B}=\hbar =1$).  \cite{graphite}
%$\thinspace $\rho _{ab}$
Extending the upper limit of the temperature range
to 900 K,
we observe an unexpected
pronounced increase in the in-plane resistivity.
We explain this effect by
% accounting for
interaction of carriers with high-frequency, graphene-like optical
phonons.

%We begin with a brief description of the graphite spectrum in the relevant
%energy range and ensuing expectations for the temperature dependence of $%
%\rho _{ab}.$
The band structure of graphite
can be characterized by three
energy scales. \cite{graphite} The largest one is set by the hopping matrix element between
A and B atoms in graphene sheets: $\gamma _{0}\approx 3.2$ eV. The
next two largest ones
%is
comprise
the matrix elements between
% nearest neighbors in
adjacent
Bernal-stacked graphene sheets: $\gamma _{1}\approx 0.35$ eV (for the vertical bond) and $\gamma_3\approx 0.3$ eV (for the slanted bond). Next
comes a large number of matrix elements between next-to-nearest neighbors,
which are known to less accuracy, but are generally believed not to exceed
several tens of meV. The major difference between graphite and graphene,
i.e., an overlap of the conduction and valence band which gives rise to a
small ($\approx 3\times 10^{18}$ cm$^{-3}$ at $T\rightarrow 0$) but finite
carrier concentration,
% and, hence, finite Fermi energy, comes about because of
is due to hopping between next-to-nearest planes. The band overlap and the Fermi
energy are on the order of this matrix element.
%: the Fermi energy is about $%
%E_{F}=250 $ K \cite{graphite}(we set $k_{B}=\hbar =1)$.
%Most of the
 Prior
%experimental
transport measurements of graphite were performed at or below $300$ K, where it
behaves as a compensated semi-metal with finite Fermi energy. There is,
however, a very interesting but hitherto unexplored regime of temperatures $%
E_{F}\lesssim T\lesssim \gamma _{1}\approx 4060$ K.
This regime
can be viewed as a  critical region of the quantum phase transition between
a semi-metal with finite band overlap and a semiconductor with finite band gap.
%In this regime,
%only two energy scales ($%
%\gamma _{0} $ and $\gamma _{1}$) matter, while all
%all low-energy couplings
Since next-to-nearest plane couplings are irrelevant in this regime, graphite can be thought of as a stack of graphene
bi-layers.
\cite{comment_graphene} We dub this regime as \lq\lq bi-layer
graphite\rq\rq (BLGT).
%Due to a drastic reduction in the number of relevant
%couplings,
If slanted hopping ($\gamma_3$) is also neglected, BLGT
%can be
is described by a simple (and historically the first)
Wallace model of graphite, \cite{Wallace} which contains only two hoppings: $\gamma _{0} $ and $\gamma _{1}$. The energy
spectrum in this model consists of two electron and two hole
branches %\begin{equation}
%\varepsilon _{\mathbf{k}}=\pm \gamma _{1}\Gamma \pm \sqrt{\gamma
%_{1}^{2}\Gamma ^{2}+\hbar ^{2}v_{0}^{2}k_{||}^{2}},  \label{eq:wallace1}
%\end{equation}
\begin{equation}
\varepsilon _{\mathbf{k}}=\pm \gamma _{1}\Gamma \pm \sqrt{\gamma
_{1}^{2}\Gamma ^{2}+\gamma _{0}^{2}|S_{\mathbf{k}}|^{2}},
\label{eq:wallace1}
\end{equation}
where $\Gamma =\cos \left( k_{z}c/2\right) ,$ $c$ is the $c$-axis lattice
constant and $S_{\bf{k}}$ is
the
structure factor for hopping between inequivalent ($A$ and
$B)$ atoms:
% depends on the in-plane components of
%the momentum
%$\bf{k}$:
\begin{equation}
S_{\mathbf{k}}=e^{ik_{x}a/\sqrt{3}}+2e^{-ik_{x}a/2\sqrt{3}}\cos (ak_{y}/2).
\end{equation}
Near the $K$ and $K^{\prime }$ points of the Brillouin zone, $\gamma _{0}S_{%
\mathbf{k}}\approx v_{0}\xi _{\mathbf{k}}\equiv v_{0}\left(
k_{x}+ik_{y}\right) $, where $v_{0}$ is the Dirac velocity of a single
graphene layer. The density of states in BLGT is energy-independent
%\begin{equation}
%DM
$\nu =16\gamma _{1}/v_{0}^{2}c$
%,  \label{DOS}
%\end{equation}
up to $\mathcal{O}(\varepsilon /\gamma _{1})$ terms (Ref.~\onlinecite{Wallace}).
Consequently,
the number density of charge carriers
 increases linearly with $T$
%Because of the linear increase of
%the density with $T$,
%Consequently,
and the in-plane  conductivity scales
linearly with $T\tau$ (Ref.~\onlinecite{Wallace})
\begin{equation}
\sigma _{ab}=\left(4\ln 2/\pi\right)\left(e^{2}/c\right)T\tau.  \label{sigma_wallace}
\end{equation}
%again up to $\mathcal{O}(T/\gamma _{1})$.
At high temperatures, when scattering is predominantly due
to phonons, one expects $1/\tau $ to scale linearly with $T$ and,
consequently,  $\sigma _{ab}$ to be $T$ independent. The
conductivity measured previously up to $300$K does indeed show a tendency to saturation, \cite{graphite} in accord with
this expectation. However, a different behavior is observed for
$T>300$K.

\section{Experiment}

\begin{figure}[t]
\includegraphics[angle=0,width=0.5\textwidth]{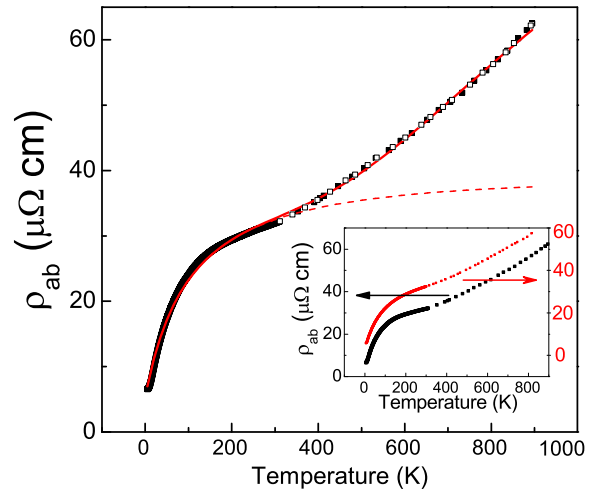}
\caption{(Color online). Measured temperature dependence
%t
of the in-plane resistivity of HOPG for warming (filled squares) and cooling (blank squares) temperature sweeps. Dashed:
% compared with
theoretical prediction for $\rho_{ab}(T)$ in the model containing scattering at impurities and soft phonons.
%scattering rate
%$1/\tau=1/\tau_0+\alpha T$ and realistic graphite spectrum.
% (solid red curve)
Solid: fit using the model containing scattering at impurities, soft phonons, and intervalley scattering at hard in-plane optical phonons [Eq. (\ref{fit})].
% (solid green curve). The
Inset:
%shows
overlap of the data for two
%separate
different samples. The vertical scales were shifted  for clarity.}
\label{fig1}
\end{figure}

\begin{figure}[t]
\includegraphics[angle=0,width=0.5\textwidth]{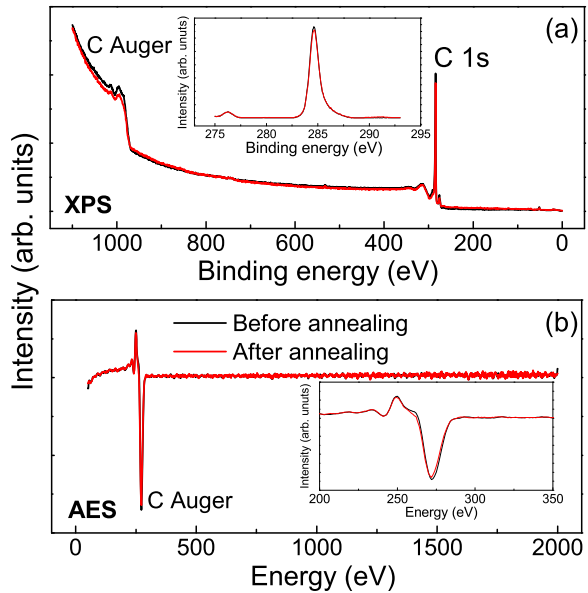}
\caption{ (Color online). (a) X-ray
photoelectron spectroscopy (XPS) spectra before and after annealing. (b) Auger electron spectroscopy (AES)  spectra before and after annealing. Insets: magnified carbon peaks.}
\label{fig2}
\end{figure}

We
%DM
%have
measured the in-plane resistivity of highly oriented pyrolytic graphite (HOPG) from $290$ K up to $900$ K in a sealed oven and from $1.7$ K up to $310$ K in Physical Property Measurement System (PPMS) using LR700 $16$ Hz AC resistance bridge in four terminal contact configuration. The samples
%have been
were
cleaved before measurements. Reproducibility of the results was checked by superimposing the temperature-dependent resistivities of two separate samples (Fig.~\ref{fig1} inset). All HOPG samples were identified to have 0.5 degree mosaic spread determined by X-ray rocking curve measurements. Gold contacts to the sample were made using silver or graphite paint. Reproducibility of the measurements was checked by sweeping the temperature up and down in the range $1.7~\mathrm{K}\leq T\leq 900$ K.  High temperature measurements were performed in a sealed oven with nitrogen flowing gas. Temperature was recorded by a type J thermocouple located $\sim 1$mm from the sample to minimize temperature lagging effects. Each data point was taken after reaching temperature stability. The room temperature resistivity was measured
to be $\rho_{ab}^{300\mathrm{K}}\approx 32\;\mu\Omega\cdot$cm before and after annealing, consistent with the existing literature values. \cite{graphite} Since $\rho_{ab}^{300\mathrm{K}}$ remains at the same value before and after annealing, we conclude that adsorption/desorption of impurities or doping
%has not taken
does not occur
%place
up to 900~K. In addition, X-ray photoelectron spectroscopy (XPS, Fig.~\ref{fig2}a) and Auger electron spectroscopy (AES, Fig.~\ref{fig2}b)
%have been
were performed on different samples and contamination was not observed.
The XPS and AES spectra displayed respectively
a characteristic C1s peak at 284.6 eV and a
C auger peak at 271.8 eV. The shape and position of the XPS C1s peak was unaltered indicating that C remained in the same chemical state.
When relating the measured resistance to resistivity, the
lattice constant was assumed to be constant. We estimate the change of the c-axis lattice constant to be 1\%
at the highest temperature measured. \cite{Steward,Mounet}
%, which corresponds only to a 3\% increase in $\gamma_1$ and may be neglected.

The experimental results for $\rho_{ab}(T)$  are presented in
Fig.~\ref{fig1} as points. Note that the tendency
to saturation pronounced at $T\sim 300$~K is superseded by a rapid
increase which continues unabated up to the highest temperature measured.
The dashed line in the top panel shows the theoretical
prediction for $\rho _{ab}\left( T\right)$, calculated for $1/\tau =1/\tau
_{0}+\alpha T$ and for a realistic energy spectrum of carriers.
While this model describes the experiment at low temperatures,
it fails completely for $T>300$ K. A slow increase in the theoretical curve,
which amounts only for a $\sim 12\%$
increase of $\rho_{ab}$ from $300$ to $900$ K,
is due to corrections of order $T/\gamma _{1}$ to
%DM
%the Wallace result,
Eq.~(\ref{sigma_wallace}).
%over the range of temperatures

\section{Theory}

To explain the observed $T$-dependence of $\rho_{ab}$, we first
notice  that the highest resistivity measured,
$\rho_{ab}^{895\mathrm{K}}\approx 62.5\;\mu
\Omega\cdot \mathrm{cm}$, corresponds to  $2~\mathrm{k}\Omega$  per graphene
sheet, i.e., well below the resistance quantum, $h/e^2\approx 25\; \mathrm{k}\Omega$.
Therefore, the Boltzmann equation  should provide an adequate description of
transport in the entire temperature interval. To explain the data,
we thus need to invoke a new scattering mechanism,
with $1/\tau $ increasing faster than $T$.

\subsection{Qualitative picture of transport}

Before proceeding  with a discussion of such  mechanisms, it
is instructive to develop a more intuitive picture of transport in BLGT. \
To this end,
%we notice that the
we replace the Wallace spectrum
%can be reduced to
by a massive
(Galilean) form obtained by expanding Eq.~(\ref{eq:wallace1}) in $k_{||}\equiv \sqrt{%
k_{x}^{2}+k_{y}^{2}}$ and keeping only the degenerate branches of electrons
and holes:
\begin{equation}
\varepsilon _{\mathbf{k}}^{\pm }=\pm
%\frac{
k_{||}^{2}/
%}{
2m_{||}\left(
k_{z}\right)
%}
,  \label{galilean}
\end{equation}
where $m_{||}\left( k_{z}\right) =\Gamma \gamma _{1}/v_{0}^{2}$ is the $%
k_{z} $-dependent mass of the in-plane motion. Evaluating the in-plane conductivity as
\begin{equation}
\sigma
_{ab}\left( T\right) =4e^{2}\int d^{3}k\left( -\partial f^{0}
%_{T}
_{\bf{k}}/\partial
\varepsilon _{\mathbf{k}}\right) v_{||}^{2}\tau \left( \varepsilon _{\mathbf{%
k}},T\right) /(2\pi )^{3}\label{sigma_ab}
\end{equation} with spectrum given by Eq.~(\ref{galilean}), $\tau=\mathrm{const}$, and $f^{0}_{\mathbf{k}}=1/\left(\exp\left(\varepsilon_k/T\right)+1\right)$, we reproduce Eq.~(\ref{sigma_wallace}).
% is given by
%\begin{equation}
%\sigma _{ab}=\frac{e^{2}\tau }{\pi ^{2}}\int d\omega \left( -f_{T}^{\prime
%}\right) \int_{-\pi /c}^{\pi /c}\frac{dk_{z}}{2\pi }\int dk_{||}^{2}\delta
%\left( \varepsilon _{k}^{+}-\omega \right) \frac{k_{||}^{2}}{\left[
%m_{||}\left( k_{z}\right) \right] ^{2}}\,  \label{boltzmann}
%\end{equation}
%which reproduces Eq.~(\ref{sigma_wallace}).
Typical momenta contributing to
$\sigma_{ab}$ are $k_{z}\sim 1/c$ and $k_{||}\sim k_{T}\equiv \sqrt{2%
\bar{m}_{||}T},$ where $\bar{m}_{||}\equiv m_{||}\left( k_{z}=0\right)
=\gamma _{1}/v_{0}^{2}.$ Although expansion in $k_{||}$ breaks down near the
$H$ points ($k_{z}=\pm \pi /c$), where $\Gamma$ vanishes and the spectrum
is Dirac-like, the contribution of Dirac fermions to $\sigma _{ab}$ is small
in proportion to the volume of the Brillouin zone they occupy.
%DM
% and
%leads to  corrections of order $\mathcal{O}(\varepsilon /\gamma_{1})$
%to Eq.~(\ref{sigma_wallace})
\cite{comment_caxis}
Therefore, a typical
carrier (in zero magnetic field) in BLGT is massive rather than Dirac-like and the isoenergetic surfaces
are corrugated cylinders  centered near the $K$ points.
%and the density of states is energy-independent.
As in the case of bi-layer graphene,  \cite{trigonal}  $\gamma_3$ hopping (responsible for trigonal warping of the isoenergetic surfaces)
leads to a linear-in-$k_{||}$ term in the energy spectrum, which is smaller than the quadratic term for energies
 $>\gamma_1\gamma_3^2/\gamma_0^2\sim 30$ K. Since we are not interested here in special effects arising solely from trigonal warping, e.g., longitudinal magnetoresistance,
 this term can be safely neglected.

Having this simple picture in mind, we now turn to a discussion of various
scattering mechanisms.
\subsection{Scattering mechanisms}

\subsubsection{Electron-hole interaction}

We start with the \textit{electron-hole} interaction. In contrast to the
electron-electron interaction, this mechanism gives rise to finite
resistivity in a compensated semi-metal even in the absence of Umklapp
processes. \cite{kekkonen,Levinson} For $T\ll E_{F},$ we have a usual Fermi-liquid
behavior $1/\tau _{\text{e-h}}\propto T^{2}.$ The $T^{2}$-behavior of $\rho_{ab}$ is indeed observed in graphite below $5$ K. \cite{old_ee,Du}
However, BLGT is not a Fermi liquid but rather a non-degenerate
electron-hole plasma with fixed (and equal to zero) chemical potential. To
estimate the strength of Coulomb interaction, we calculate the $r_{s}$
parameter, i.e. the ratio of typical potential and kinetic energies
\begin{equation}
r_{s}(T)=e^{2}/\epsilon_{\infty} lE_{\mathrm{kin}},
\end{equation}
where $\epsilon_{\infty} $ is the background dielectric constant of graphite and $l=(4\pi
/3)^{1/3}n^{-1/3}\propto T^{-1/3}$ is the typical inter-carrier distance.
Using an experimentally measured values of number density $n\approx 10^{19}\;$cm$%
^{-3}$ at $T=300$ K (Ref.~\onlinecite{us_unpublished}) and $\epsilon_{\infty}\approx 5$ (Ref.~\onlinecite{epsilon})
and evaluating the kinetic energy as $E_{\mathrm{kin}}=\left( \pi
^{2}/12\ln 2\right) T\approx 1.2T$ (Ref. \onlinecite{comment_partition}), we
obtain $r_{s}(T=300$ K$)\approx 1.2$. As temperature increases, $r_{s}\left(
T\right) $ decreases as $T^{-2/3}$. Therefore, the perturbation theory for
the electron-hole interaction is  reasonably accurate already at $T\simeq 300$
K  and becomes even better at higher $T$. In the Thomas-Fermi model, the screened Coulomb potential $U\left(
\mathbf{q}\right) =4\pi e^{2}/\epsilon_{\infty} \left( q_{||}^{2}+q_{z}^{2}+\kappa
^{2}\right)$ is isotropic even if the electron spectrum is not; all details of the spectrum are encapsulated in the (square of) screening wavenumber $\kappa ^{2}=4\pi e^{2}\nu /
\epsilon_{\infty} $  proportional to the density of states. In BLGT, $\kappa^2$ is $T$-independent and, 
since $\nu $ is
proportional to the (small) in-plane mass, $\kappa \ll c^{-1}$. Also, at not too high
temperatures, $k_{T}\propto \sqrt{T}\ll \kappa$. Since $q_{||}$
cannot exceed the typical electron momentum, $q_{||}\lesssim k_{T}\ll
\kappa $ and $U\left( q\right) $ almost does not depend on $q_{||}$: $%
U\left( \mathbf{q}\right) \approx 4\pi e^{2}/\epsilon_{\infty} \left(
q_{z}^{2}+\kappa ^{2}\right) .$ \ Using this form of $U\left( q\right) $ and
the spectrum from Eq.~(\ref{galilean}), we obtain from the Fermi Golden
Rule
\begin{equation}
\frac{1}{\tau _{\text{e-h}}}\sim \frac{e^{2}\gamma _{1}}{\epsilon \kappa
v_{0}^{2}}T\,.
\end{equation}
This linear scaling persists
even at higher temperatures ($T\gtrsim \gamma_1$), in the single-layer limit. \cite{guinea_ee}
Therefore,  electron-hole interaction does not provide an explanation of the experiment.
\subsubsection{Electron-phonon interaction}

We now turn to the \textit{electron-phonon} scattering.
With 4 atoms per
unit cell, graphite has 3 acoustic and 9 optical phonon modes. The
phonon spectrum consists of two groups of modes: \lq\lq hard\rq\rq and \lq\lq soft\rq\rq.
 \cite{wirtz} The characteristic energy scale of hard modes, which are present already in graphene, is $\sim 0.1$ eV.
%corresponds to Debye energies of the in-plane acoustic phonons and
%frequencies of the in-plane optical phonons with in-phase displacements of
%atoms in adjacent
%graphene
%sheets.
Soft modes, with typical energies of
order $\ 10$ meV, arise from weak coupling between
graphene
sheets.
%as well
%as from out-of-phase displacements of atoms in adjacent sheets.
For the
temperatures of interest  ($T>300$ K), all soft modes are in the
classical regime, in which the occupation number and, thus, the
scattering rate scale linearly with $T$.  Although hard acoustic modes are
still below their Debye temperatures, they are also in the classical regime.
For example, typical in-plane phonon momenta involved in scattering at a
hard, graphene-like acoustic mode with dispersion $\omega _{A}=s_{ab}q_{||}$
are $\bar{q}_{||}\sim \bar{k}_{||}$. The corresponding frequencies $\bar{%
\omega}_{A}\sim s_{ab}\bar{k}_{||}$ are smaller than $T$ as long as $T\gg
\bar{m}_{||}s_{ab}^{2}\sim 1$ K. Therefore, all soft modes as well as  hard acoustic
modes lead to linear scaling of the scattering rate
%\begin{equation*}
$\tau _{\mathrm{e-ph}}^{-1}=\alpha T$.
%\end{equation*}
%Within the deformation potential model, the dimensionless coefficient is
%small due to a small value of the in-plane effective mass; according to a
%recent experiment, $\alpha \approx 0.065.$
%Switching briefly to another scattering mechanism--electron-electron interaction--we notice that although $1/\tau_{\mathrm{e-e}}$ due to this mechanism scales as $T^2$ for $T<E_F$,
%this scaling does not hold at higher $T$. Indeed, the screened Coulomb potential in BLGT does not practically depend on the-in-plane momentum transfer and can be regarded as a constant, while the number of carriers scales linearly with $T$.
%Consequently, $1/\tau_{\mathrm{e-e}}\propto T$. 

The remaining hard modes are graphene-like optical phonons,
%i.e.,
e.g.,
the
longitudinal optical ($E_{2\text{g}})$ mode with frequency $\ \omega
_{0}\approx 0.17$ eV at the $\Gamma $ point. \cite{wirtz,graphite_raman}
For $T\lesssim \omega _{0},$ scattering at these modes leads to an
exponential growth of the resistivity with temperature. In the remainder of
the paper, we will show that this mechanism is capable of explaining the
experiment.

Even if only hard phonons are taken into account, a real picture of the
electron-phonon interaction in graphite is rather complicated, as both
inter- and intra-valley scattering
% due
on a number of modes are involved. Since
our goal is just to obtain the temperature dependence of the resistivity, we
will construct a simplified model, \cite{ando,antonio} taking into account only inter-valley
scattering due to one optical mode. We start from a tight-binding  Hamiltonian
\begin{equation}
H_{\mathrm{graphite}}=
\begin{pmatrix}
H_{\parallel } & H_{\perp } \\
H_{\perp }^{\dagger } & H^{{*}}_{\parallel }
\end{pmatrix}
\!.  \label{H_graphite}
\end{equation}
Here, $H_{\parallel }$ describes hopping within graphene sheets
\begin{equation}
H_{\parallel }=
\begin{pmatrix}
\gamma _{0}^{\prime }\tilde{S}_{\mathbf{k}} & \gamma _{0}S_{\mathbf{k}} \\
\gamma _{0}S_{\mathbf{k}}^{\ast } & \gamma _{0}^{\prime }\tilde{S}_{\mathbf{k%
}}
\end{pmatrix}
,
\end{equation}
where $\tilde{S}_{\mathbf{k}}=4\cos \left( \sqrt{3}k_{x}a/2\right) \cos \left(
k_{y}a/2\right) +2\cos \left( k_{y}a\right) $ is the structure factor for
in-plane hopping between next-to-nearest neighbors ($AA$ and $BB$). As we will show shortly,
%show
% the dominant  part of
the electron-phonon interaction
is dominated
%arises from
by the coupling between  the diagonal part of
$H_{\parallel }$
%to
and phonons.
%DM
%where we keep only the largest out-of-plane matrix element, $\gamma _{1}.$
Near the $K$ and $K^{\prime }$ points, $\tilde{S}\approx
-3+3k_{||}^{2}a^{2}/4.$ The momentum-dependent part of $\tilde{S}$ changes
the spectrum only at large ($\sim 1/a$) \ $k_{||}$ and, therefore, can be
neglected. Hopping in the $c$-direction is described by
\begin{equation}
H_{\perp }=
\begin{pmatrix}
2\gamma _{1}\cos (k_{z}c/2) & 0 \\
0 & 0
\end{pmatrix}
\!.
\label{graphite}\end{equation}
%Up to an additive constant, the eigenvalues $%
%\varepsilon _{\mathbf{k}}$
The spectrum of Hamiltonian (\ref{H_graphite}) is given by Eq.(\ref
{eq:wallace1}), while the eigenvectors are represented by a spinor
%\begin{equation}
$\lambda _{\mathbf{k}}=C\left( \pm \varepsilon _{\mathbf{k}}/\xi _{%
\mathbf{k}},\pm 1,\varepsilon _{\mathbf{k}}/\xi _{\mathbf{k}}%
,1\right)$
%\end{equation}
near the $K$ point and by $\lambda _{\mathbf{k}}^{\ast }$ near the $K^{\prime }$
point. The elements of $\lambda _{\mathbf{k}}$ are the amplitudes of finding
a charge carrier on one of the four atoms $(A,B,\tilde{A},\tilde{B})$ of the
graphite unit cell.
%DM commenting out
It is important to notice a difference between $\lambda
_{\mathbf{k}}$ and the graphene spinors $\mu _{\mathbf{k}}=\left( 1,\pm
\varepsilon _{\mathbf{k}}/\xi _{\mathbf{k}}\right) .$
Near the $K$ points, $%
\varepsilon _{\mathbf{k}}=|\xi _{\mathbf{k}}|$ in graphene and, therefore, $A
$ and $B$ atoms are occupied with equal probabilities.
If the optical phonon
frequency is
%still
much smaller than
%the
$\gamma _{1}$
% hopping element
,
 both
the initial and final states of a scattering process are described by the
spectrum in Eq.~(\ref{galilean}). For this spectrum,  the amplitudes on $A$
and $\tilde{A}$ atoms are small as $\left| \varepsilon _{\mathbf{k}}\right|
/\left| \xi _{\mathbf{k}}\right| \sim \bar{k}_{||}/\bar{m}_{||}v_{0}\sim
\left( T/\gamma _{1}\right) ^{1/2}\ll 1$. \cite{antonio}  Neglecting these amplitudes, we
replace the eigenvectors by $\lambda _{\mathbf{k}}^{0}=\left( 0,\pm
1,0,1\right) /\sqrt{2}.$ In reality, $\omega _{0}/\gamma _{1}\approx 0.5$
and corrections to the results following from this approximation would be
important in a more detailed theory.

In the simplest model, phonons modulate hopping amplitudes by stretching the
corresponding bonds. The Hamiltonian of this interaction has the same
structure as that for an ideal lattice
\begin{equation}
H_{\mathrm{e-ph}}=
\begin{pmatrix}
\tilde{H}_{\parallel } & \tilde{H}_{\perp } \\
\tilde{H}_{\perp }^{\dagger } & \tilde{H}_{\parallel }^{*}
\end{pmatrix}
\!,
\end{equation}
where tilde denote corrections to hopping matrix elements due to lattice
distortions. If $\tilde{H}_{\parallel }$ does not have diagonal elements, \
the matrix element of $H_{\mathrm{e-ph}}$ between the spinors $\lambda _{%
\mathbf{k}}^{0}$ vanishes. Therefore, the electron-phonon interaction
appears only in the diagonal elements, describing modulation of hopping
between the next-to-nearest neighbors. Expanding Hamiltonian (\ref
{H_graphite}) in atomic displacements, one finds
\begin{eqnarray}
&&\!\!\!H_{\mathrm{e-ph}}^{AA}=i\frac{\partial \gamma _{0}^{\prime }}{%
\partial a}\sum_{q}\,\,\,\mathbf{F\cdot \hat{u}}_{A,\mathbf{q}}\,\,  \notag
\\
F_{x} &=&\sin aq_{x}+2\sin
%\frac{
\left(aq_{x}/
%}{
2\right)
%}
\cos\left(
%\frac{
\sqrt{3}
%}{
%}
aq_{y}/2\right)\,
\notag \\
&&F_{y}=2\sqrt{3}\sin
\left(
% \frac{
a\sqrt{3}q_{y}/2\right)\cos \left(aq_{x}/2\right)\,,
\label{electron_phonon}
\end{eqnarray}
where $\mathbf{\hat{u}}_{A,\mathbf{q}}$ is the displacement operator
for\ $A$ atoms in momentum space. Displacements in the $c$-direction,
being perpendicular to the plane, do not change (to linear order) the
distance between adjacent $A$ atoms, and are therefore uncoupled from
fermions. Since the formfactors $F_{x}$ and $F_{y}$ are small for small $q,$
processes with larger $q$ have higher probabilities. For this reason, we will
focus on inter-valley scattering between $K$ and $K^{\prime }$ points,
corresponding to $\mathbf{q}=\mathbf{q}_{0}\equiv 2\pi /a\left( -1/\sqrt{3}%
,1/3\right) .$

Further expansion in displacements generates higher-order vertices of the
electron-phonon interaction which  account for anharmonic effects.
%\DM{, such
%as multiple phonon  emission/absorption.}
  Although anharmonicity is important in a monolayer graphene,\cite{Eros} it
is weak in bulk graphite.
For example, the temperature variation of the $c$-axis thermal expansion
coefficient amounts only to 6\% in the temperature interval from 273 to 1000 K. \cite{anharmonic} Therefore, we neglect the anharmonic effects
for
% the time being
now but will return to this
point when discussing
%the comparison with
the experiment.
Employing Eq.~(\ref{electron_phonon}), we find the matrix element
%\begin{equation}
$M_{\mathbf{k}_{1}\rightarrow \mathbf{k}_{2}}^{\mathbf{q}}=\delta _{\mathbf{k}%
_{2},\mathbf{k}_{1}+\mathbf{q}}\Upsilon _{\mathbf{q}}/\sqrt{2\rho
\omega _{\mathbf{q}}L^{3}}\,$,
%\end{equation}}
%\begin{equation}
%M_{\mathbf{k}_{1}\rightarrow \mathbf{k}_{2}}^{\mathbf{q}}=\delta _{\mathbf{k}%
%_{2},\mathbf{k}_{1}+\mathbf{q}}\frac{1}{L^{3/2}}\sqrt{\frac{\hbar }{2\rho
%\omega _{\mathbf{q}}}}\Upsilon _{\mathbf{q}}\,,
%\end{equation}
where $\Upsilon _{\mathbf{q}}=\left( \partial \gamma _{0}^{\prime }/\partial
a\right) \mathbf{d_{q}}\cdot\mathbf{F}$ is the deformation
potential for optical phonons, $\mathbf{d_{q}}$ is the
% dimensionless
polarization vector, $L$ is a system size, and $\rho $ is the mass density.
%$\xi_{\bf q}=\sqrt{1/2\rho\omega_{\bf q}}$ is the amplitude
%of zero point motion in the mode ${\bf q}$.
The corresponding scattering time can be evaluated as \cite{Levinson}
\begin{equation}
\tau _{\text{e-ph}}^{-1}(\mathbf{k}_{1})=\sum_{k_{2}}
%W_{\mathbf{k}%_{1}\rightarrow \mathbf{k}_{2}}^{\mathbf{q}}
\left(W_{\mathbf{k%
}_{1}\rightarrow \mathbf{k}_{2}}^{+\mathbf{q}}+W_{\mathbf{k}_{1}\rightarrow
\mathbf{k}_{2}}^{-\mathbf{q}}\right)
%\frac{
\left(1-f^{0}
%_{T}(
_{\mathbf{k}_{2}}
%)
\right)
%}{%
\left(1-f^{0}
%_{T}(
_{\mathbf{k}_{1}}
)\right)^{-1}
%}
\,,
\end{equation}
where
%$W_{\mathbf{k}_{1}\rightarrow \mathbf{k}_{2}}^{\mathbf{q}}=W_{\mathbf{k%
%}_{1}\rightarrow \mathbf{k}_{2}}^{+\mathbf{q}}+W_{\mathbf{k}_{1}\rightarrow
%\mathbf{k}_{2}}^{-\mathbf{q}}$
% is the total scattering rate,
%and $W^{\pm {\bf q}}$
%are
the transition rates of emission and absorption of phonons,
respectively, are
\begin{equation}
W_{\mathbf{k}_{1}\rightarrow \mathbf{k}_{2}}^{\pm \mathbf{q}}\!\!=2\pi |M_{%
\mathbf{k}_{1}\rightarrow \mathbf{k}_{2}}^{\mathbf{q}}|^{2}\!\!\left( N_{%
\mathbf{q}}\!+\!\frac{1}{2}\pm \frac{1}{2}\right) \delta \left(
\!\varepsilon _{\mathbf{k}_{1}}\!-\!\varepsilon _{\mathbf{k}_{2}}\mp \omega
_{\mathbf{q}}\!\right) ,
\end{equation}
and $N_{\mathbf{q}}$ is the Bose function.
Neglecting the dispersion of the optical mode and using the simplified
electron spectrum from Eq.~(\ref{galilean}), we find for the inter-valley
scattering rate
\begin{equation}
\tau _{\text{iv}}^{-1}(\varepsilon _{\mathbf{k}},T)=\bar{\tau}^{-1}
\frac{\coth\left(\omega_0/2T\right)\cosh^2\left(\varepsilon_{\mathbf{k}}/2T\right)}{\cosh^2\left(\varepsilon_{\mathbf{k}}/2T\right)+\sinh^2\left(\omega_0/2T\right)},
%\frac{(1+e^{\varepsilon _{\mathbf{k}}/T})^{2}\coth \left( \frac{\omega _{0}}{%
%2T}\right) }{\left( 1+e^{(\varepsilon _{\mathbf{k}}+\omega _{0})/T}\right),
%\left( 1+e^{(\varepsilon _{\mathbf{k}}-\omega _{0})/T}\right) }\,,
\end{equation}
where all model-dependent details of the electron-phonon interaction are
incorporated into
%a nominal scattering rate
$\bar{\tau}^{-1}$.
%Within our
%model,
%%\begin{equation}
%\bar{\tau}^{-1}=\frac{3m_{\parallel }}{16\pi \rho \omega _{0}}\left( \frac{%
%\partial \gamma _{0}^{\prime }}{\partial a}\right) ^{2}\int_{-\pi /c}^{\pi
%/c}dq_{z}\mathbf{d}_{\mathbf{q}}^{2}(\mathbf{q}_{0},q_{z}).
%\end{equation}
For a thermal electron ($\varepsilon _{\mathbf{k}}\sim T$)
the scattering rate behaves as $\exp \left( -\omega _{0}/T\right)$
for $T\ll \omega _{0}$. For $T\gg
\omega _{0}$,  scattering at this mode crosses over into the classical regime and $\tau
_{\text{iv}}^{-1}\sim T/\omega_0\bar{\tau}$.

At sufficiently high temperatures,
% ($T\gg\omega _{0}$),
% the
where intervalley scattering is the
dominant mechanism,
%and
the conductivity is obtained by substituting
$\tau _{\text{iv}}(\varepsilon _{\mathbf{k}},T)$ into Eq.~(\ref{sigma_ab})
\begin{equation}
%\sigma _{ab}^{\left( \text{iv}\right) }=\frac{\ln 2-1/4}{3\pi ^{2}}\frac{%
%e^{2}}{c}T\tilde{\tau}\ln \left( \frac{\gamma _{1}}{T}\right) \exp \left(
%\frac{\omega _{0}}{T}\right) ,  \label{sigma_iv}
\sigma _{ab}^{\left( \text{iv}\right) }=\left[\left(4\ln 2-1\right)/3\pi\right]
\left(e^{2}/c\right)T\bar{\tau}\exp \left(\omega _{0}/T\right)\; ,  \label{sigma_iv}
\end{equation}
%where $\tilde{\tau}$ is the nominal time $\bar{\tau}$ evaluated at $m_{||}=%
%\bar{m}_{||}.$ A logarithmic factor arises because $\bar{\tau}$, being
%proportional to $1/m_{||},$ vanishes at the $H$ points and the $k_{z}$
%integral diverges logarithmically. Notice that a small ($\approx 0.015$)
%numerical prefactor in Eq.~(\ref{sigma_iv}) makes the intervalley
%contribution to the resistivity significant even for $T\ll \omega _{0}$.
\subsection{Comparison to experiment}
Based on the result for the conductivity for intervalley scattering [Eq.~(\ref{sigma_iv})], we fit the observed $\rho _{ab}(T)$ by
the following formula
\begin{equation}
\rho _{ab}=
%\frac{\pi }{4\ln 2}
\frac{c}{e^{2}}
\left( \frac{1}{\tau _{0}}%
+\alpha T\right) \frac{1}{\varepsilon ^{\ast }}+\frac{c}{e^{2}}\frac{1}{%
a_0T\bar{\tau}
%\ln \left( \gamma _{1}/T\right)
}Ä
\exp \left( -\frac{\omega _{0}%
}{T}\right),
\label{fit}
\end{equation}
where $a_0=2\left(4\ln2-1\right)/3\pi\approx 0.376$ and $\varepsilon ^{\ast }\equiv c \int
d^{3}kv_{||}^{2}\left( -\partial f^{0}_{\bf{k}}/\partial \varepsilon _{\mathbf{k}%
}\right)/2\pi^3$.
The first term in Eq.~(\ref{fit}) accounts mostly for the
low-$T$ behavior of $\rho _{ab},$ when scattering at impurities ($1/\tau _{0}
$) and soft phonons ($\alpha T$) dominate transport. When calculating $\varepsilon ^{\ast}$,
we accounted for hopping between next-to-nearest planes, described by  $\gamma_{2}\approx 0.02$ eV.
For finite $\gamma _{2}$, the Fermi energy is finite, $\epsilon^*\sim E_F$ at $T\to 0$, and the first term in Eq. ~(\ref{fit}) goes to a finite value at $T\to 0$.  The second term is due
to intervalley scattering.
%DM
%The coefficient in the first term was
%chosen to match Eq.(\ref{sigma_wallace}).
Equation (%
\ref{fit}) contains four fitting parameters: $\tau _{0},\alpha ,\bar{\tau}$, and $\omega _{0.}$ The results of the fit are shown in Fig.~\ref{fig1}.  The fitting parameters are $\tau_0=6.29\times10^{-12}$ s,
$\alpha=0.09$, $\bar{\tau}=1.4\times 10^{-14}$ s, and $\omega_0=0.22$ eV. The values of $\tau_0$ and $\alpha$ are in reasonable agreement
with those found previously. \cite{Du} The frequency $\omega_0$ is somewhat higher but still close to the frequency of the $E_{2\mathrm{g}}$ mode. \cite{wirtz,graphite_raman}
A rather short nominal time $\bar{\tau}$ indicates strong coupling between electrons and optical phonons in graphite.

%Alternative explanation
For completeness, we note that for  $T$ above the Debye frequency,
the multi-phonon
% scattering rate is proportional to $T^n$, where
%$n$ is a number of phonons involved. This lead to
processes modify the scattering rate as
%scattering rate
$1/\tau=1/\tau_0+\alpha \left(T+(T/T_2)^2+(T/T_3)^3+{\cal O}(T^4)\right)$,
%Then, single-phonon processes contribute
%a $T$ term to $1/\tau$, two-phonon processes contribute a  $T^2$ term, etc.
%The meaning of $T_{2,3}$ is the temperature at which two(three)-phonon
%scattering becomes comparable to single-phonon one. Physically,
where $T_{2,3}$
correspond to an energy scale at which anharmonicity becomes strong.
The data can be fitted by  $T_2\approx T_3\approx 10^{3}$ K, which
is well below the scale of $10^{4}$ K at which anharmonicity becomes strong in
other physical properties, such as the $c$-axis thermal expansion coefficient
%$\alpha_c(T)$
 and elastic moduli. \cite{anharmonic} This reinforces our conclusion
that anharmonicity is not important in transport for $T<1000$K.
%For example, extrapolating the measured $T$ dependence of $\alpha_c(T)$ to high%er temperatures \cite{anharmonic}, we find
%that $\alpha_c(T)$ changes by 100\% only at $T\sim 10^{4}$ K.
\section{Conclusions}
To summarize, we have studied, both experimentally and theoretically,
the in-plane resistivity of HOPG.  We found that its temperature dependence
is determined by a competition between those of the carrier number density, $n(T)$,  and
of the scattering rate $1/\tau$.   At temperatures below $50$ K, the number density is practically
independent of the temperature, while the the scattering rate increases with the temperature; as a result, the resistivity
increases with $T$. At temperatures comparable to the Fermi energy, the increase in $n(T)$ almost compensates for that in $1/\tau$,
leading to a quasi-saturation of $\rho_{ab}$ at $T\sim 300$ K. However, full saturation never occurs because,
as the temperature increases further,  scattering off hard optical phonons,
characterized by an exponential increase of $1/\tau$ with $T$, becomes important.
This results in a further increase of $\rho_{ab}$ with $T$.
\acknowledgments
We thank D. Aristov, D. Bagrets, R. Bowers, H.-P. Cheng, I. Dmitriev, J.-N. Fuchs, M. Goerbig, I. Gornyi, E. Lambers, N. Kirova, A. Mirlin, P.
Ostrovky, D. Polyakov,  C. Stanton, A. Shnirman, and S. Trickey for interesting discussions. The HOPG
samples were supplied by J. Fischer (UPenn). D.B.G. acknowledges hospitality
of the Department of Physics of MUN. D.L.M. acknowledges the financial support from RTRA Triangle de
la Physique and hospitality of the Laboratoire de Physique des Solides,
Universit{\'e} Paris-Sud. A.F.H acknowledges support by NSF under Grant No. 0704240.


\begin{thebibliography}{99}
\bibitem{comment_graphene}
At energies larger than $\gamma_1$,  accessible in optical experiments,
graphite
%fully decouples into graphene layers.
can be viewed as a stack of decoupled graphene layers.
\bibitem{comment_partition}  The equipartition theorem of the classical
statistical physics, which states that the kinetic energy associated with
every degree of freedom is equal to $T/2,$ is not applicable to a
compensated semimetal (even if the spectrum massive), because the
distribution function is not Maxwellian but rather a Fermi-Dirac one with
zero chemical potential.
\end{thebibliography}
\end{document}